\documentclass[fleqn,10pt]{wlscirep}
\usepackage[utf8]{inputenc}
\usepackage[T1]{fontenc}
\usepackage{lineno}
\usepackage{amsmath,amssymb,amsthm,amsfonts,epsfig,siunitx}
\usepackage{xcolor}
\usepackage{etoolbox}
\usepackage{caption,multirow,hyperref,pifont,graphicx,enumitem,colortbl}
\usepackage[frozencache]{minted}
\usepackage{mdframed}
\DeclareCaptionType{Listing}[Box]
\hypersetup{breaklinks=true,colorlinks=true,citecolor=blue,filecolor=blue,linkcolor=blue,urlcolor=blue}

\newcommand{\onlinecite}[1]{\nocite{#1}\citenum{#1}}

\newcommand{\listingref}[1]{Box \ref{#1}}
\newcommand{\webref}[1]{\href{#1}{#1}}


\usepackage{minted, mdframed}
\setminted[json]{
    numbers=left,
    breaklines,
    style=emacs,
    breakbefore=\&/?=,
    fontsize=\footnotesize, 
    linenos
}

\mdfdefinestyle{mintframe}{
    innertopmargin=2pt,
    innerbottommargin=2pt,
    innerleftmargin=2pt,
    innerrightmargin=2pt,
    backgroundcolor=blue!3,
    skipbelow=0,
    skipabove=0,
    linecolor=gray!50,
}

\usepackage[scaled=0.9]{inconsolata}
\usepackage{listings}

\definecolor{eclipseStrings}{RGB}{42,0.0,255}
\definecolor{eclipseKeywords}{RGB}{127,0,85}

\definecolor{darkpastelblue}{rgb}{0.47, 0.62, 0.8}
\definecolor{darkpastelgreen}{rgb}{0.01, 0.75, 0.24}
\definecolor{darkpastelpurple}{rgb}{0.59, 0.44, 0.84}
\definecolor{darkorchid}{rgb}{0.6, 0.2, 0.8}
\definecolor{darkpink}{rgb}{0.91, 0.33, 0.5}
\definecolor{deepskyblue}{rgb}{0.0, 0.75, 1.0}
\definecolor{dodgerblue}{rgb}{0.12, 0.56, 1.0}
\definecolor{dogwoodrose}{rgb}{0.84, 0.09, 0.41}
\definecolor{dollarbill}{rgb}{0.52, 0.73, 0.4}
\colorlet{numb}{darkpastelgreen}
\colorlet{keywords}{dodgerblue}
\colorlet{strings}{dogwoodrose}
\colorlet{punct}{black}
\lstdefinelanguage{json}{
    basicstyle=\footnotesize\ttfamily,
    commentstyle=\color{strings},
    stringstyle=\color{keywords},
    numberstyle=\ttfamily,
    stepnumber=1,
    numbersep=8pt,
    upquote=true,
    showstringspaces=false,
    breaklines=true,
    frame=single,
    backgroundcolor=\color{blue!2},
    string=[b]",
    morestring=[b]",
    literate=
        *{0}{{{\color{numb}0}}}{1}
         {1}{{{\color{numb}1}}}{1}
         {2}{{{\color{numb}2}}}{1}
         {3}{{{\color{numb}3}}}{1}
         {4}{{{\color{numb}4}}}{1}
         {5}{{{\color{numb}5}}}{1}
         {6}{{{\color{numb}6}}}{1}
         {7}{{{\color{numb}7}}}{1}
         {8}{{{\color{numb}8}}}{1}
         {9}{{{\color{numb}9}}}{1}
         {.}{{{\color{numb}.}}}{1}
         {:}{{{\color{punct}{:}}}}{1}
         {,}{{{\color{punct}{,}}}}{1}
}
\lstdefinelanguage{html}{
    basicstyle=\footnotesize\ttfamily,
    stringstyle=\color{strings},
    commentstyle=\color{keywords},
    stepnumber=1,
    numbersep=8pt,
    showstringspaces=false,
    breaklines=true,
    string=[s]{"}{"},
    comment=[l]{:\ "},
    morecomment=[l]{:"}
}
\title{\begin{center}\includegraphics{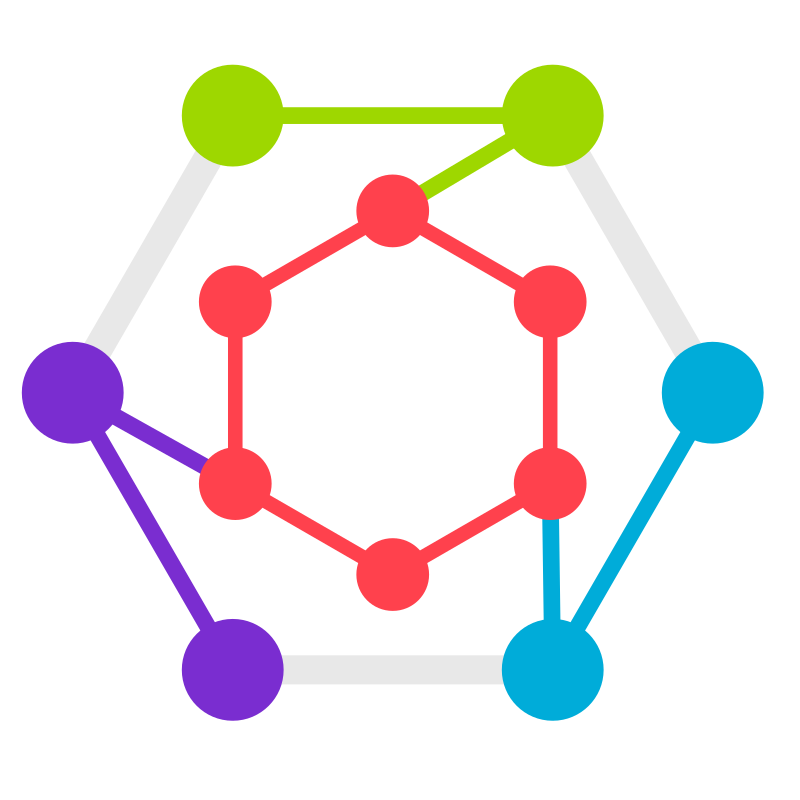}\end{center}
OPTIMADE, an API for exchanging materials data}
\author[1,$\dag$]{Casper W. Andersen}
\author[2,$\dag$]{Rickard Armiento}
\author[3,4,$\dag$]{Evgeny Blokhin}
\author[5,$\dag$]{Gareth J. Conduit}
\author[6,$\dag$]{Shyam Dwaraknath}
\author[5,7,$\dag$]{Matthew L. Evans}
\author[8,7,$\dag$]{Ádám Fekete}
\author[9,$\dag$]{Abhijith Gopakumar}
\author[10,11,$\dag$]{Saulius Gra\v{z}ulis}
\author[10,$\dag$]{Andrius Merkys}
\author[12,$\dag$]{Fawzi Mohamed}
\author[13,14,$\dag$]{Corey Oses}
\author[1,$\dag$]{Giovanni Pizzi}
\author[7,$\dag$]{Gian-Marco Rignanese}
\author[12,15,$\dag$]{Markus Scheidgen}
\author[1,16,$\dag$]{Leopold Talirz}
\author[13,14,$\dag$]{Cormac Toher}
\author[6,$\dag$]{Donald Winston}
\author[17,18]{Rossella Aversa}
\author[19]{Kamal Choudhary}
\author[13,14]{Pauline Colinet}
\author[13,14]{Stefano Curtarolo}
\author[20]{Davide Di Stefano}
\author[15]{Claudia Draxl}
\author[21]{Suleyman Er}
\author[13,14]{Marco Esters}
\author[22,13]{Marco Fornari}
\author[7]{Matteo Giantomassi}
\author[23]{Marco Govoni}
\author[7]{Geoffroy Hautier}
\author[9]{Vinay Hegde}
\author[6]{Matthew K. Horton}
\author[6]{Patrick Huck}
\author[15]{Georg Huhs}
\author[24]{Jens Hummelshøj}
\author[25]{Ankit Kariryaa}
\author[26,27]{Boris Kozinsky}
\author[1]{Snehal Kumbhar}
\author[9]{Mohan Liu}
\author[1]{Nicola Marzari}
\author[28]{Andrew J. Morris}
\author[29]{Arash A. Mostofi}
\author[6,30]{Kristin A. Persson}
\author[7]{Guido Petretto}
\author[12]{Thomas Purcell}
\author[7]{Francesco Ricci}
\author[13,14]{Frisco Rose}
\author[12]{Matthias Scheffler}
\author[12,15]{Daniel Speckhard}
\author[1]{Martin Uhrin}
\author[10]{Antanas Vaitkus}
\author[4]{Pierre Villars}
\author[7]{David Waroquiers}
\author[9]{Chris Wolverton}
\author[6]{Michael Wu}
\author[31]{Xiaoyu Yang}
\affil[1]{Theory and Simulation of Materials (THEOS), and National Centre for Computational Design and Discovery of Novel Materials (MARVEL), \'{E}cole Polytechnique F\'{e}d\'{e}rale de Lausanne, 1015 Lausanne, Switzerland}
\affil[2]{Materials Design and Informatics unit, Department of Physics, Chemistry and Biology, Link\"oping University, Sweden}
\affil[3]{Tilde Materials Informatics, Stra{\ss}mannstra{\ss}e 25, 10249, Berlin, Germany}
\affil[4]{Materials Platform for Data Science, Sepapaja 6, 15551, Tallinn, Estonia}
\affil[5]{Theory of Condensed Matter Group, Cavendish Laboratory, 19 J.J. Thomson Avenue, Cambridge, CB3 0HE, United Kingdom}
\affil[6]{Lawrence Berkeley Labs, Berkeley, CA, USA}
\affil[7]{UCLouvain, Institut de la Mati\`ere Condens\'ee et des Nanosciences (IMCN), Chemin des \'Etoiles~8, Louvain-la-Neuve 1348, Belgium}
\affil[8]{Department of Physics, King's College London, London, United Kingdom}
\affil[9]{Department of Materials Science and Engineering, Northwestern University, Evanston, IL 60208, USA}
\affil[10]{Institute of Biotechnology, Life Science Center, Vilnius University, Saul\'{e}tekio av.\ 7, LT-10257 Vilnius, Lithuania}
\affil[11]{Institute of Computer Science, Faculty of Mathematics and Informatics, Vilnius University, Naugarduko g. 24, LT-03225 Vilnius, Lithuania}
\affil[12]{Fritz-Haber-Institut der Max-Planck-Gesellschaft, Faradayweg 4-6, 14195, Berlin, Germany}
\affil[13]{Center for Autonomous Materials Design, Duke University, Durham, NC 27708, USA}
\affil[14]{Department of Mechanical Engineering and Materials Science, Duke University, Durham, NC 27708, USA}
\affil[15]{Humboldt-Universit\"at zu Berlin, Institut f\"ur Physik and IRIS Adlershof, 12489 Berlin, Germany}
\affil[16]{Laboratory of Molecular Simulation (LSMO), \'Ecole Polytechnique F\'ed\'erale de Lausanne, 1951 Sion, Switzerland}
\affil[17]{National Research Council-Istituto Officina dei Materiali (CNR-IOM), 34136, Trieste, Italy}
\affil[18]{Karlsruhe Institute of Technology (KIT), Hermann-von-Helmholtz-Platz 1, 76344 Eggenstein-Leopoldshafen, Germany}
\affil[19]{Materials Measurement Laboratory, National Institute of Standards and Technology, Gaithersburg, MD, 20899, USA}
\affil[20]{Ansys, 300 Rustat House, 62 Clifton Rd, Cambridge CB1 7EG, United Kingdom}
\affil[21]{Dutch Institute for Fundamental Energy Research (DIFFER), De Zaale 20, 5612 AJ, Eindhoven, The Netherlands}
\affil[22]{Department of Physics and Science of Advanced Materials Program, Central Michigan University, Mount Pleasant, Michigan 48859, USA}
\affil[23]{Argonne National Laboratory, Lemont, IL 60439, USA}
\affil[24]{Thayer School of Engineering, Dartmouth College, Hanover, NH, 03755, USA}
\affil[25]{Toyota Research Institute (TRI), Los Altos, California 94022, USA}
\affil[26]{Department of Computer Science, University of Copenhagen, Universitetsparken 1, 2100 Copenhagen, Denmark}
\affil[27]{John A. Paulson School of Engineering and Applied Sciences, Harvard University, Cambridge, Massachusetts 02138, USA}
\affil[28]{Robert Bosch LLC, Research and Technology Center North America, 255 Main St, Cambridge, Massachusetts 02142, USA}
\affil[29]{School of Metallurgy and Materials, University of Birmingham, Edgbaston, Birmingham, B15 2TT, United Kingdom}
\affil[30]{Departments of Materials and Physics, and the Thomas Young Centre, Imperial College London, Exhibition Road, London SW7 2AZ, United Kingdom}
\affil[31]{Department of Material Science and Engineering, Hearst Mining Memorial Building, UC Berkeley, Berkeley CA 94720, USA.}
\affil[32]{Computer Network Information Center,  Chinese Academy of Sciences, Beijing, China}
\affil[33]{University of Chinese Academy of Sciences, Beijing, China}
\affil[$\dag$]{These authors contributed equally to this work, and can be contacted at dev@optimade.org}

\begin{abstract} 
The Open Databases Integration for Materials Design (OPTIMADE) consortium has designed a universal application programming interface (API) to make materials databases accessible and interoperable. We outline the first stable release of the specification, v1.0, which is already supported by many leading databases and several software packages. We illustrate the advantages of the OPTIMADE API through worked examples on each of the public materials databases that support the full API specification.
\end{abstract}

\begin{document}

\flushbottom
\maketitle

\thispagestyle{empty}

\section*{Introduction}

Data has become a crucial resource in many scientific fields, and materials science is no exception.
Experimental data has long been meticulously curated in handbooks and databases, with the first edition of Landolt-B\"ornstein\cite{Landolt-Bornstein} being published in 1883.
Nowadays, various commercial and non-commercial experimental databases, such as the Inorganic Crystal Structure Database (ICSD)~\cite{Zagorac2019}, are widely used throughout the field.

High-throughput electronic structure calculations, themselves enabled by algorithmic improvements and growing computational resources, have significantly increased the availability of useful data from computational simulations of materials. Since the pioneering work of Ceder~\textit{et al.}\cite{Ceder1998}, a large number of high-throughput first-principles studies have been reported in the literature (for a review, see Ref.~\onlinecite{Curtarolo2013}), with results typically collated in databases. This explosion in the amount of available data has kick-started a new paradigm of data-driven materials science\cite{Himanen2019}, creating opportunities for concurrent, automated materials design, boosted by databases that can be queried by humans and machines via an application programming interface (API)\cite{AFLOW_REST_API,AFLUX,MP_REST_API,MaterialsCloud}.

As materials databases differ in fidelity and focus across material classes and properties, it is extremely beneficial to be able to liberate and unify data from multiple sources. However, retrieving data from multiple databases is difficult as each database has its own specialized, and sometimes esoteric, API that governs data access patterns, querying and the representation of the underlying data. Moreover, as the APIs of individual databases inevitably evolve, existing clients must also evolve; a significant maintenance effort is required to translate the responses from the new API to the representation of the client.

Motivated by these considerations, providers of several materials databases united to design and implement an API specification that enables seamless access and interoperability across materials databases. The effort started at the workshop ``Open Databases Integration for Materials Design'', held at the Lorentz Center in Leiden, Netherlands in October 2016, and continued at followup workshops held at CECAM in Lausanne, Switzerland in June 2018, June 2019, and June 2020. The result is the OPTIMADE specification (v1.0)\cite{OPTIMADE2020}; OPTIMADE defines a RESTful API that is queried with URLs, with responses adhering to the JSON:API specification\cite{jsonapi}. Specification development adheres to Semantic Versioning~\cite{SemVer} to avoid surprises and enable backwards-compatibility where possible, without impeding further development. By extracting the technical and scientific commonalities from existing APIs, the OPTIMADE API has been designed so that it can be implemented across a broad range of materials domains, database back-ends and sizes.

In this paper, we first review the query format of existing databases to motivate the design and construction of the OPTIMADE API specification. We then illustrate the use of the API with a set of worked examples;  databases that already fully support the OPTIMADE API are enumerated alongside their results for representative queries in Table~\ref{tab:optimade_endpoints}. We further highlight libraries that could accelerate uptake and assist materials data curators to support the OPTIMADE API format. Finally, we discuss future prospects and ongoing development of the OPTIMADE API.

\section*{Current generation of materials database APIs}

Materials databases are a veritable treasure trove of information, but they only become useful once a human, or machine, can access them. In this section we review the current range of APIs used by various databases to enable access to an example compound, SiO$_2$, which serves to highlight the variation of APIs that a user must navigate in order to make use of multiple materials databases. We then demonstrate the universal nature of the OPTIMADE API that permits seamless access to all materials databases that support it.

We first compare and contrast the APIs that must be used to request records on an exemplar system, SiO$_2$, from three different databases: AFLOW, the Materials Project, and the Crystallographic Open Database (COD). All three queried databases support requests using a representational state transfer through a web service (RESTful), at the following URLs:
\begin{description}[labelwidth=2.8cm]
 \item[AFLOW] {\small\url{http://aflow.org/API/aflux/?species(Si,O),nspecies(2)}}
 \item[Materials Project] {\small\url{https://www.materialsproject.org/rest/v2/materials/SiO2/vasp/structure}}
 \item[COD] {\small\url{https://www.crystallography.net/cod/result.php?formula=O2\%20Si}}
\end{description}
Note that the Materials Project requires the user to supply an API key (\webref{http://materialsproject.org/open}) preferably specified in the \texttt{X-API-KEY} HTTP header.

The three APIs vary syntactically (in format), taxonomically (having different names for terms), and semantically (in the conflicting definitions of chemical formula as an intensive or extensive property). AFLOW returns all structures with both Si and O present, whereas both the Materials Project and COD deliver any structure with a formula unit of SiO$_2$. The wide range of query formats that will deliver non-overlapping structures significantly complicates access to all available data for SiO$_2$, without even considering the differing representations of the structures returned.

The inconsistent format of the query is further complicated by the difficulty of accessing other structures with the SiO$_2$ formula. Focusing on just AFLOW, two possible queries that users more familiar with the other APIs might attempt are
\begin{description}
  \item{\small{\url{http://aflow.org/API/aflux/?compound(SiO2)}}} \hfill \\
 which returns no response;
 \item{\small{\url{http://aflow.org/API/aflux/?compound(O2Si1)}}} \hfill \\
 now lists the elements in alphabetical order as required by AFLOW, and includes the ``1'' after element symbols, so that ``\texttt{SiO2}'' becomes ``\texttt{O2Si1}''. This returns entries where the unit cell is SiO$_2$, but does not return Si$_2$O$_4$ or simulation cells containing more formula units.
 \end{description}
The exemplar {\small{\url{http://aflow.org/API/aflux/?species(Si,O),nspecies(2)}}} returns all entries with at least one Si and an O, so while the response includes the SiO$_2$ phases of interest, it may also contain other stoichiometries.

The distinctions between the request format for each database require the user to become an expert in many different APIs. This again emphasises the need for a single well-designed and standardized API to access all materials databases, which is the aim of the OPTIMADE API.

\section*{The OPTIMADE API}

The OPTIMADE API provides a holistic standard for serving and accessing the information in compatible materials databases. To retrieve information about materials from a particular database, the user submits a request via a URL. Each database provider will have published a base URL that serves the OPTIMADE API, for example {\verb|https://example.com/optimade/|}. The same URL path, across different OPTIMADE API implementations, allows uniform access to the underlying databases. Both human-readable and machine-readable versions of the OPTIMADE API specification are available online with releases archived at Zenodo \cite{OPTIMADE2020}. The specification is also registered as an API standard on FAIRsharing.org\cite{fairsharing}.

\subsection*{Design philosophy}

The OPTIMADE specification strives to enable materials information to be filtered and retrieved in a straightforward and intuitive manner. The three queries from the previous section can each be performed on a standardised, versioned endpoint ({\color{OliveGreen}{\verb|/v1/structures|}}) that enables access to a \verb|structures| entry resource type that consists of many well-defined attributes. The specification then defines a grammar for filtering entries against these attributes, allowing the previous SiO$_2$ filter example to be expressed in a common way ({\color{darkorchid}{\verb|?filter=chemical_formula_reduced="O2Si"|}}). Altogether, the universal OPTIMADE URL, where only the implementation URL changes, becomes:
\begin{center}
\begin{minipage}{0.75\textwidth}
{\texttt{{\color{blue}<optimade\_implementation\_url>}{\color{OliveGreen}/v1/structures\color{darkorchid}?filter=chemical\_formula\_reduced="O2Si"}}}
\end{minipage}
\end{center}

The OPTIMADE specification also aims to be flexible to many different underlying data representations, and thus there are very few properties that are mandatory. Instead of enforcing an exhaustive set of property definitions, individual OPTIMADE implementations can describe the data they serve via {\color{OliveGreen}\verb|/info|} endpoints for each entry type. These introspective endpoints allow clients to adapt to the particular implementation for an underlying database, and allow providers to disseminate properties beyond the simple structural and chemical information standardized by the specification. To avoid naming collisions, each provider-specific property name must be prefixed by a provider-specific token, itself bookended by underscores (\texttt{\_}). The property \texttt{custom\_property} from an example provider with assigned prefix \texttt{exmpl} would be expressed as \texttt{\_exmpl\_custom\_property}: e.g., \texttt{\_tcod\_a} for the lattice constant, $a$, in TCOD, or \texttt{\_aflow\_spacegroup\_relax} for the space group of the relaxed structure in AFLOW.

\subsection*{Implementation discovery}

The list of implementations confirmed and tested in this paper to support the OPTIMADE API is shown in Table~\ref{tab:optimade_endpoints}. They are all publicly accessible, providing users with open access to large international repositories of computational and experimental materials science data.

The OPTIMADE consortium provides an open, federated list of implementations  (\webref{https://providers.optimade.org}). It is considered to be a catalogue of currently available and/or known public OPTIMADE implementations. New implementations are welcome and can register themselves via a pull request on GitHub (\webref{https://github.com/Materials-Consortia/providers}).

The requirements for appearing in the above providers list are very loose. Some databases listed in the catalogue are signalling the intent of future implementations, while others only have partial implementations of the OPTIMADE API, including JARVIS (\webref{https://jarvis.nist.gov/optimade}) \cite{choudhary2020jarvis} and MatCloud (\webref{https://matcloud.com.cn}) \cite{Yang2018}. Some software frameworks, such as AiiDA \cite{AiiDA,AiiDA2,AiiDA3}, also enable users to access their personal data through an OPTIMADE API, and therefore have a dedicated provider-specific ID, but no single official OPTIMADE implementation base URL.

The OPTIMADE API also specifies an endpoint for semi-automated cross-provider discovery. The \texttt{/links} endpoint serves \texttt{links} resources that may refer to either provider internal (\texttt{child}, \texttt{root}) or external (\texttt{external}, \texttt{providers}) resources based on the \texttt{link\_type} attribute. To avoid being overly restrictive, it is at the provider's discretion whether they serve a list of known providers; however, this provides a mechanism for scalable and decentralised discovery of new implementations beyond the federated provider list.

\begin{table}[ht]
\begin{center}
\footnotesize
\newcolumntype{a}{>{\columncolor{blue!3}}l}
\newcolumntype{b}{>{\columncolor{blue!3}}c}
\newcolumntype{d}{>{\columncolor{blue!3}}r}
\renewcommand{\arraystretch}{1.1}
\begin{tabular}{|a|ddd|}
\hline
\bf{Provider}&\bf{N$_1$}&\bf{N$_2$}&\bf{N$_3$}\\\hline
\href{http://www.aflow.org}{AFLOW}\cite{AFLOW_database, aflow_fleet_chapter} & \href{http://aflow.org/API/optimade/v1/structures?filter=elements HAS ANY "C","Si","Ge","Sn","Pb"}{700,192} & \href{http://aflow.org/API/optimade/v1/structures?filter=elements HAS ANY "C","Si","Ge","Sn","Pb" AND nelements=2}{62,293} & \href{http://aflow.org/API/optimade/v1/structures?filter=elements HAS ANY "C","Si","Ge","Sn" AND NOT elements HAS "Pb" AND elements LENGTH 3}{382,554}\\
\href{https://www.crystallography.net/cod}{Crystallography Open Database} (COD)\cite{Grazulis_COD_2009, Grazulis_COD_2012} & \href{https://www.crystallography.net/cod/optimade/v1/structures?filter=elements HAS ANY "C","Si","Ge","Sn","Pb"}{416,314} & \href{https://www.crystallography.net/cod/optimade/v1/structures?filter=elements HAS ANY "C","Si","Ge","Sn","Pb" AND nelements=2}{3,896} & \href{https://www.crystallography.net/cod/optimade/v1/structures?filter=elements HAS ANY "C","Si","Ge","Sn" AND NOT elements HAS "Pb" AND elements LENGTH 3}{32,420}\\
\href{https://www.crystallography.net/tcod}{Theoretical Crystallography Open Database} (TCOD)\cite{Merkys_TCOD_2017} & \href{https://www.crystallography.net/tcod/optimade/v1/structures?filter=elements HAS ANY "C","Si","Ge","Sn","Pb"}{2,631} & \href{https://www.crystallography.net/tcod/optimade/v1/structures?filter=elements HAS ANY "C","Si","Ge","Sn","Pb" AND nelements=2}{296} & \href{https://www.crystallography.net/tcod/optimade/v1/structures?filter=elements HAS ANY "C","Si","Ge","Sn" AND NOT elements HAS "Pb" AND elements LENGTH 3}{660}\\
\href{https://materialscloud.org}{Materials Cloud}\cite{AiiDA,AiiDA2,MaterialsCloud} & \href{https://aiida.materialscloud.org/optimade-sample/optimade/v1/structures?filter=elements HAS ANY "C","Si","Ge","Sn","Pb"}{886,518} & \href{https://aiida.materialscloud.org/optimade-sample/optimade/v1/structures?filter=elements HAS ANY "C","Si","Ge","Sn","Pb" AND nelements=2}{801,382} & \href{https://aiida.materialscloud.org/optimade-sample/optimade/v1/structures?filter=elements HAS ANY "C","Si","Ge","Sn" AND NOT elements HAS "Pb" AND elements LENGTH 3}{103,075}\\
\href{http://materialsproject.org}{Materials Project}\cite{Materials_Project,Jain_2011, Ong_pymatgen_2013,Mathew_Atomate_CMS_2017} & \href{https://optimade.materialsproject.org/v1/structures?filter=elements HAS ANY "C","Si","Ge","Sn","Pb"}{27,309} & \href{https://optimade.materialsproject.org/v1/structures?filter=elements HAS ANY "C","Si","Ge","Sn","Pb" AND nelements=2}{3,545} & \href{https://optimade.materialsproject.org/v1/structures?filter=elements HAS ANY "C","Si","Ge","Sn" AND NOT elements HAS "Pb" AND elements LENGTH 3}{10,501}\\
\href{https://nomad-lab.eu}{Novel Materials Discovery Laboratory} (NOMAD)\cite{NOMAD_2017,NOMAD_2018} & \href{https://nomad-lab.eu/prod/rae/optimade/v1/structures?filter=elements HAS ANY "C","Si","Ge","Sn","Pb"}{3,359,594} & \href{https://nomad-lab.eu/prod/rae/optimade/v1/structures?filter=elements HAS ANY "C","Si","Ge","Sn","Pb" AND nelements=2}{532,123} & \href{https://nomad-lab.eu/prod/rae/optimade/v1/structures?filter=elements HAS ANY "C","Si","Ge","Sn" AND NOT elements HAS "Pb" AND elements LENGTH 3}{1,611,302}\\
\href{https://odbx.science}{Open Database of Xtals} (odbx)\cite{odbx-matador} & \href{https://optimade.odbx.science/v1/structures?filter=elements HAS ANY "C","Si","Ge","Sn","Pb"}{55} & \href{https://optimade.odbx.science/v1/structures?filter=elements HAS ANY "C","Si","Ge","Sn","Pb" AND nelements=2}{54} & \href{https://optimade.odbx.science/v1/structures?filter=elements HAS ANY "C","Si","Ge","Sn" AND NOT elements HAS "Pb" AND elements LENGTH 3}{0}\\
\href{http://openmaterialsdb.se}{Open Materials Database} (\textit{omdb})\cite{HTTKOMDB} & \href{http://optimade.openmaterialsdb.se/v1/structures?filter=elements HAS ANY "C","Si","Ge","Sn","Pb"}{58,718} &
\href{http://optimade.openmaterialsdb.se/v1/structures?filter=elements HAS ANY "C","Si","Ge","Sn","Pb" AND nelements=2}{690} & \href{http://optimade.openmaterialsdb.se/v1/structures?filter=elements HAS ANY "C","Si","Ge","Sn" AND NOT elements HAS "Pb" AND elements LENGTH 3}{7,428}\\
\href{http://oqmd.org}{Open Quantum Materials Database} (OQMD)\cite{OQMD} & \href{http://oqmd.org/optimade/v1/structures?filter=elements HAS ANY "C","Si","Ge","Sn","Pb"}{153,113} & \href{http://oqmd.org/optimade/v1/structures?filter=elements HAS ANY "C","Si","Ge","Sn","Pb" AND nelements=2}{11,011} & \href{http://oqmd.org/optimade/v1/structures?filter=elements HAS ANY "C","Si","Ge","Sn" AND NOT elements HAS "Pb" AND elements LENGTH 3}{70,252}\\
\hline
\end{tabular}
\end{center}
\caption{Materials databases with active OPTIMADE API implementations and the number of entries they return for the filters presented in this paper. The OPTIMADE website provides an up-to-date record of the implementation status of the databases (\webref{https://www.optimade.org/providers-dashboard/}).
AFLOW, Materials Project, odbx, \textit{omdb}, and OQMD comprise computational materials data generated using database-specific workflows \cite{aflow_fleet_chapter,Materials_Project, odbx-matador, HTTKOMDB, OQMD}.
For the purposes of this table, Materials Cloud results were aggregated across all provided sub-databases.
COD and TCOD comprise experimental and theoretical crystal structure data extracted from the literature.
Materials Cloud comprises materials data from computational workflows; sub-databases group data by research project and can be contributed by users \cite{MaterialsCloud}.
NOMAD aggregates computational data from multiple sources including from several of the repositories listed here. }
\label{tab:optimade_endpoints}
\end{table}

\section*{Worked example}

To illustrate the effective use of the OPTIMADE API we now provide a worked example of querying structures. We explore materials containing Group 14 elements (the carbon family), starting with a general search before drilling down to specific materials. The Group 14 elements are of particular interest as their atomic orbitals regularly hybridise, enabling a variety of bonding with differing geometries. The hybridised orbitals enable these elements to form the backbone of a wide range of compounds, both inorganic and organic, that underpin plastics, drugs, and semiconductors. Group 14 therefore forms both a diverse and important family of compounds that heavily populates databases, so are an ideal case study to demonstrate the OPTIMADE API.

\subsection*{Common features of the response}

Whilst our previous exploration of the Group 14 compounds considered only SiO$_2$, the flexibility of the OPTIMADE API allows us to start with a search over all materials in Group 14, comprising carbon (C), silicon (Si), germanium (Ge), tin (Sn), and lead (Pb). We start with a simple API call that searches for all materials that contain at least one element in Group 14:
\begin{center}
\begin{minipage}{0.6\textwidth}
{\texttt{{\color{OliveGreen}/v1/structures}{\color{darkorchid}?filter=elements HAS ANY "C", "Si", "Ge", "Sn", "Pb"}}}
\end{minipage}
\end{center}
This string can be appended to the base URL of any of the available implementations, to gather results in a standardised form.
The base URL can be found on the providers dashboard (\webref{https://www.optimade.org/providers-dashboard}).

As an example, this query is run through the Theoretical Crystallography Open Database (TCOD)\cite{Merkys_TCOD_2017} with the following URL:
\begin{center}
\href{https://www.crystallography.net/tcod/optimade/v1/structures?filter=elements HAS ANY "C","Si","Ge","Sn","Pb"}{\texttt{{\color{blue}https://www.crystallography.net/tcod/optimade}{\color{OliveGreen}/v1/structures}{\color{darkorchid}?filter=elements HAS ANY  "C", "Si", "Ge", "Sn", "Pb"}}}
\end{center}
The JSON response is summarized in Boxes \ref{listing:attributes} through \ref{listing:meta}, where some lines have been omitted for brevity; the full response is given in Supplementary File 1.

The first tranche of the JSON response comprises the ``\texttt{data}'' field that contains a list of entries returned for the query; a truncated version of this field is shown in \listingref{listing:attributes}, displaying a few salient properties of just one of the ten entries from the full response. The response for a particular material entry comprises multiple sections:

\begin{Listing}[h!]
\begin{center}
\fbox{
\begin{minipage}{0.55\textwidth}
\begin{mdframed}[style=mintframe]
\begin{minted}{json}
{
  "data": [
    {
      "attributes": {
\end{minted}
\end{mdframed}
\begin{mdframed}[style=mintframe]
\begin{minted}{json}
        "dimension_types": [1, 1, 1],
        "elements": ["O", "Sn", "Ta"],
\end{minted}
\end{mdframed}
\begin{mdframed}[style=mintframe]
\begin{minted}{json}
        "lattice_vectors": [
          [4.02510400822, 0, 0],
          [0, 4.02510400822, 0],
          [0, 0, 4.02510400822]
        ],
\end{minted}
\end{mdframed}
\begin{mdframed}[style=mintframe]
\begin{minted}{json}
        "_tcod_a": 4.02510400822,
        "_tcod_alpha": 90,
        "_tcod_b": 4.02510400822,
        "_tcod_beta": 90,
        "_tcod_c": 4.02510400822,
\end{minted}
\end{mdframed}
\end{minipage}
}
\end{center}
  \caption{An excerpt of the JSON response showing the material attributes for one of the returned \texttt{structures} entries.\label{listing:attributes}}
\end{Listing}

\begin{description}
\item[\texttt{attributes}]\listingref{listing:attributes} shows the physical properties of the material comprising both mandatory information such as \texttt{elements} and \texttt{lattice\_vectors}, as well as optional, additional database-specific information prefixed with the database name (e.g., \texttt{\_tcod\_}, here used to provide lattice parameters). This ensures that all databases return the most important and common information in a standardized format, as well as allowing them to include additional database-specific data. Importantly, the OPTIMADE specification provides a standardized way for database implementations to be self-documenting, via introspective \texttt{/info} endpoints. We see in the \texttt{elements} section that here we have returned a material comprising the element of interest, Sn, as well as O and Ta.
\end{description}

\begin{Listing}[h!]
\begin{center}
\fbox{
\begin{minipage}{0.55\textwidth}
\begin{mdframed}[style=mintframe]
\begin{minted}{json}
      "id": "10000003",
      "links": {
        "self": "http://www.crystallography.net/tcod/10000003.cif"
      },
\end{minted}
\end{mdframed}
\end{minipage}
}
\end{center}
  \caption{An excerpt of the JSON response showing the top-level entry ID and a self-link.\label{listing:id_links}}
\end{Listing}

\begin{description}
\item[\texttt{id} and \texttt{links}]\listingref{listing:id_links} shows the unique ID for the entry within the database, and a self-link to the database-specific representation/rendering of the entry (in this case, a link to a Crystallographic Information File\cite{cif_2016}).
\end{description}

\begin{Listing}[h!]
\begin{center}
\fbox{
\begin{minipage}{0.55\textwidth}
\begin{mdframed}[style=mintframe]
\begin{minted}{json}
      "relationships": {
        "references": {
          "data": {
            "id": "10000003",
            "type": "references"
          }
        }
      }
\end{minted}
\end{mdframed}
\end{minipage}
}
\end{center}
  \caption{An excerpt JSON data response for links to other entries related to this structure in the database (here, a bibliographic \texttt{references} entry).\label{listing:relationships}}
\end{Listing}

\begin{description}
\item[\texttt{relationships}]The relationships section in \listingref{listing:relationships} links the user to other entries in the database and beyond, here the bibliographic references.
\end{description}

The additional nine materials not shown here all comprised of compounds containing either C, Si, Ge, Sn, or Pb, supplemented by a variety of other elements. The foot of the response contains information about the underlying database, comprised of three sections:

\begin{Listing}[h!]
\begin{center}
\fbox{
\begin{minipage}{0.5\textwidth}
\begin{mdframed}[style=mintframe]
\begin{minted}{json}
  "links": {
    "base_url": "http://www.crystallography.net/tcod/optimade/v1.0.0/",
    "first": {
      "href": "http://www.crystallography.net/tcod/optimade/v1.0.0/structures?page_limit=10&filter=elements HAS ANY \"C\",\"Si\",\"Ge\",\"Sn\",\"Pb\""
    },
    "next": {
      "href": "http://www.crystallography.net/tcod/optimade/v1.0.0/structures?page_limit=10&page_offset=10&filter=elements HAS ANY \"C\",\"Si\",\"Ge\",\"Sn\",\"Pb\""
    }
  },
\end{minted}
\end{mdframed}
\end{minipage}
}
\end{center}
  \caption{An excerpt of the JSON response for navigable pagination links to additional conformant materials that match the query.\label{listing:links}}
\end{Listing}

\begin{description}
\item[\ttfamily{links}]The response returned the first ten (i.e.,\ the default page limit) entries that matched the query, however, more materials are available within the database. \listingref{listing:links} shows the JSON:API-compliant pagination links to the current and next page of results for this query, as well as relevant external links.
\end{description}

\begin{Listing}[h!]
\begin{center}
\fbox{
\begin{minipage}{0.5\textwidth}
\begin{mdframed}[style=mintframe]
\begin{minted}{json}
  "meta": {
    "api_version": "1.0.0",
    "data_available": 2906,
    "data_returned": 2631,
    "implementation": {
      "maintainer": {
        "email": "cod-bugs@ibt.lt"
      },
      "name": "Theoretical Crystallography Open Database",
      "source_url": "svn://www.crystallography.net/cod/tags/tcod/v0.2.1/cgi-bin/optimade.pl@261229",
      "version": "v0.2.1"
    },
    "last_id": "10000031",
    "more_data_available": true,
    "provider": {
      "description": "Theoretical Crystallography Open Database",
      "name": "Theoretical Crystallography Open Database",
      "prefix": "tcod"
    },
    "query": {
      "representation": "/structures?filter=elements HAS ANY \"C\",\"Si\",\"Ge\",\"Sn\",\"Pb\""
    },
    "time_stamp": "2021-02-24T16:36:05Z"
  }
}
\end{minted}
\end{mdframed}
\end{minipage}
}
\end{center}
  \caption{An excerpt of the JSON response showing the metadata returned for this request.\label{listing:meta}}
\end{Listing}

\begin{description}
\item[\ttfamily{meta}]\listingref{listing:meta} provides metadata associated with the request, such as number of results, the details of the database provider, the implementation and the representation and timestamp of the submitted query.
\end{description}

A benefit of the OPTIMADE API is that the structure of the response is common to all materials databases. The responses differ only in the optional and database-specific information prefixed with the database name (here \texttt{\_tcod\_}). Table~\ref{tab:optimade_endpoints} lists several materials databases that have active OPTIMADE API implementations, and the large number of results (\textbf{N}$_1$) that they return for this particular filter.

\subsection*{Database filtering}

Requesting the example filter above from the TCOD database returns 2,631 materials entries, but the same filter could return millions (see Table~\ref{tab:optimade_endpoints}). For some requests the volume of the materials data could become unmanageable so the specification allows for the use of several pagination methods laid out by JSON:API \cite{jsonapi}. These approaches all provide a link to the ``next'' page of data:
\begin{center}
\begin{minipage}{0.85\textwidth}
{\texttt{{\color{OliveGreen}/v1/structures}{\color{darkorchid}?page\_limit=10\&page\_offset=10\&filter=elements HAS ANY "C", "Si", "Ge", "Sn", "Pb"}}}
\end{minipage}
\end{center}
with parameter \texttt{page\_offset=10} to allow the user to select which page to enter, and the parameter \texttt{page\_limit=10} to control the number of materials returned per individual request.

The most useful way to explore an OPTIMADE database is to apply a filter; the specification mandates that several relevant properties must be queryable. For example, we can perform a more focused search for materials comprising at least one element in Group 14, and a maximum of two elements (a binary material), with the filter
\begin{center}
\begin{minipage}{0.75\textwidth}
{\texttt{{\color{OliveGreen}/v1/structures}{\color{darkorchid}?filter=elements HAS ANY "C", "Si", "Ge", "Sn", "Pb" AND nelements=2}}}
\end{minipage}
\end{center}
This query returns 296 materials from the TCOD database, with the response summarized in \listingref{listing:additional}, where some lines have been omitted for brevity and the full response is given in the Supplementary File 2. The number of matching entries (\textbf{N}$_2$) across all implementations for this filter are shown in Table~\ref{tab:optimade_endpoints}.

\begin{Listing}[h!]
\begin{center}
\fbox{
\begin{minipage}{0.6\textwidth}
\begin{mdframed}[style=mintframe]
\begin{minted}{json}
{
   "data" : [
      {
         "attributes" : {
\end{minted} 
\end{mdframed}
\begin{mdframed}[style=mintframe]
\begin{minted}{json}
            "elements" : [
               "Ge",
               "O"
            ],
\end{minted} 
\end{mdframed}
\begin{mdframed}[style=mintframe]
\begin{minted}{json}
         },
\end{minted} 
\end{mdframed}
\begin{mdframed}[style=mintframe]
\begin{minted}{json}
      },
\end{minted} 
\end{mdframed}
\begin{mdframed}[style=mintframe]
\begin{minted}{json}
   ],
\end{minted} 
\end{mdframed}
\begin{mdframed}[style=mintframe]
\begin{minted}{json}
}
\end{minted} 
\end{mdframed}
\end{minipage}
}
\end{center}
  \caption{The truncated JSON response for the more focused search, showing the elements in the first material returned.\label{listing:additional}}
\end{Listing}

We can now see that the first structure, and indeed all structures, returned are comprised of at least one element in Group 14, here Ge, and a maximum of one other element (a binary material), here O. Additional filters can be chained to further refine the materials returned, or to construct more complex queries. 
For example, ternary structures that contain at least one of the elements C, Si, Ge, or Sn, but do not contain Pb (e.g., for applications where Pb toxicity would be a concern), can be retrieved using the filter
\begin{center}
\begin{minipage}{0.98\textwidth}
{\texttt{{\color{OliveGreen}/v1/structures}{\color{darkorchid}?filter=elements HAS ANY "C", "Si", "Ge", "Sn" AND NOT elements HAS "Pb" AND elements LENGTH 3}}}
\end{minipage}
\end{center}
The number of entries matching this filter are denoted as \textbf{N}$_3$ in Table~\ref{tab:optimade_endpoints}.

These simple examples demonstrate how useful chemical queries are expressible with the OPTIMADE API, allowing users to refine their queries and to suit their specific application. Further functionality of the OPTIMADE API can be found in the specification\cite{OPTIMADE2020}.

\section*{Related libraries}

The wider usage of the OPTIMADE API is a key goal for the consortium; to this end, several open source libraries have been developed to help users of the OPTIMADE API (either implementation developers, or client end-users), of which three are introduced below. The first two libraries offer tools that aid the implementation of the API for materials database developers, with the first also containing tools to construct and validate queries, while the third library is intended for end users of OPTIMADE-compliant APIs.

\subsection*{optimade-python-tools}
optimade-python-tools is an open source Python package available on GitHub (\webref{https://github.com/Materials-Consortia/optimade-python-tools}). The package contains a complete set of tools for implementing an OPTIMADE-compliant API, as well as several utilities that can be used by client code. The package is listed on the Python Package Index (PyPI) as optimade (\webref{https://pypi.org/project/optimade}). Current (v0.14) functionality of the package includes:
\begin{itemize} 
\item pydantic (\webref{https://github.com/samuelcolvin/pydantic}) data and validation models for all objects defined in the OPTIMADE specification that can be used in server or client code;
\item an extensible reference server implementation leveraging pydantic and FastAPI (\webref{https://github.com/tiangolo/fastapi}). This reference server forms the basis of the OPTIMADE implementations for the Materials Project \cite{Materials_Project}, NOMAD \cite{NOMAD_2018}, Materials Cloud\cite{MaterialsCloud} and odbx (\webref{https://odbx.science}) providers. These tools are also used to generate a machine-readable OpenAPI version\cite{openapi} of the OPTIMADE specification;
\item a validator for OPTIMADE implementations, available standalone with a command line interface (CLI) or as a GitHub action (\webref{https://github.com/Materials-Consortia/optimade-validator-action}). This validator is run against the federated list of OPTIMADE providers every day, with a live dashboard indicating compliance with the specification at \webref{https://www.optimade.org/providers-dashboard/}, the source code for which can be found on GitHub (\webref{https://github.com/Materials-Consortia/providers-dashboard});
\item a Python parser for the OPTIMADE filter language written with the Lark parsing library (\webref{https://github.com/lark-parser/lark});
\item filter transformers from the abstract syntax tree to queries to popular database back-ends. Currently, the supported back-ends are MongoDB (via pymongo) (\webref{https://github.com/mongodb}) and Elasticsearch (\webref{https://github.com/elastic/elasticsearch});
\item adapter classes to interface with other popular libraries, namely pymatgen\cite{Ong_pymatgen_2013}, ASE\cite{ase}, AiiDA\cite{AiiDA,AiiDA2,AiiDA3} and JARVIS\cite{choudhary2020jarvis} as well as converting OPTIMADE structures to the CIF\cite{cif_2016}, XYZ, and more domain standardised file formats.
\end{itemize}

\subsection*{OPTIMADE::Filter}
OPTIMADE::Filter is a Perl library for the syntactical analysis of the OPTIMADE filter language. Apart from the construction of abstract syntax trees, the library can translate simple filter strings to SQL queries. The Git repository with the source code is publicly available on GitHub (\webref{https://github.com/Materials-Consortia/OPTIMADE-Filter}).

\subsection*{pymatgen optimade module}
pymatgen\cite{Ong_pymatgen_2013} (\webref{https://pymatgen.org}) is a Python library for materials science. A user-friendly \texttt{OptimadeRester} client has been added to a new OPTIMADE module within pymatgen to provide a way to query OPTIMADE structure resources in a way familiar to existing users of pymatgen and the Materials Project API. The Git repository with the source code is publicly available on GitHub (\webref{https://github.com/materialsproject/pymatgen}).

\section*{Summary}

The latest OPTIMADE API specification v1.0\cite{OPTIMADE2020} offers holistic access to many leading crystal structure databases, namely: AFLOW, COD, TCOD, Materials Cloud, Materials Project, NOMAD, odbx, Open Materials Database (\textit{omdb}), and OQMD. 
Open client implementations are also available (\webref{https://optimade.science}, \webref{https://materialscloud.org/optimadeclient}) that enable aggregated searches over many databases as well as user-friendly graphical widgets that can create an OPTIMADE filter to empower the user with even easier access to data.
OPTIMADE provides researchers easy access to over \num[group-separator={,}]{10000000} results for different materials, providing benchmarking opportunities and offering a huge opportunity for high-throughput screening and machine learning studies. The ability of the OPTIMADE API to search databases, expose links between databases, and deliver standardized results makes it well-positioned to significantly enhance the impact and permeability of pre-existing data silos. This should empower researchers to scan through new and unexpected material families, and train models from all available data that can understand deep correlations.

The OPTIMADE API is flexible and will be extended to more use cases going forward. The development and adoption of the OPTIMADE API relies on the involvement of a large number of scientists, so contributions from the community are strongly encouraged, and questions on development, registration of a provider, or usage can be directed to the web forum (\webref{https://matsci.org/optimade}) or mailing list (\webref{dev@optimade.org}). Proposed developments include the standardization of more filterable materials properties, the integration of molecular dynamics simulations and of experimental results, and extensions beyond electronic-structure calculations. The future development of APIs, including OPTIMADE, should herald an era of effective use of big, open data in materials science.

\section*{Data availability}

The OPTIMADE specification is developed openly on GitHub with releases archived on Zenodo \cite{OPTIMADE2020}. Version 1.0.0 was released on 1 July 2020 and is licensed under Creative Commons Attribution 4.0 International (CC-BY 4.0).

\section*{Code availability}
All associated code is hosted under the Materials-Consortia organisation on GitHub (\webref{https://github.com/Materials-Consortia}).

\bibliography{main}

\section*{Acknowledgements}

The authors acknowledge support from CECAM in Lausanne (Switzerland) and the Lorentz Center in Leiden (Netherlands) for hosting OPTIMADE workshops.
CWA, SK, NM, GPi, MU, LT acknowledge financial support by the MARVEL National Centre of Competence in Research funded by the Swiss National Science Foundation (grant agreement ID 51NF40-182892), the European Centre of Excellence MaX ``Materials design at the Exascale'' (grant no. 824143) and by the the ``MARKETPLACE'' H2020 project (grant agreement ID 760173).
RAr acknowledges financial support from the Swedish e-Science Research Centre (SeRC) and the Swedish Research Council (VR) project no.\ 2016-04810 and 2020-05402.
RAv has received funding from the European Union’s Horizon 2020 Research and Innovation Program under grant agreement No. 654360 NFFA-EUROPE.
GJC would like to acknowledge financial support from the Royal Society.
SD, DWi, MWu, MH, PH, KP acknowledge financial support by the U.S. Department of Energy, Office of Science, Office of Basic Energy Sciences, Materials Sciences and Engineering Division under Contract No. DE-AC02-05-CH11231 (Materials Project program KC23MP).
CD, MLE, GH, GMR, MSchef, MSchei, and DWa acknowledge support from the European Union's Horizon 2020 research and innovation program under the European Union's Grant agreement No. 951786 (NOMAD CoE).
MLE acknowledges the EPSRC CDT in Computational Methods for Materials Science for funding under grant number EP/L015552/1.
AG, ML, VH, and CW acknowledges support by Toyota Research Institute, and the award 70NANB14H012 from U.S. Department of Commerce, National Institute of Standards and Technology as part of the Center for Hierarchical Materials Design (CHiMaD).
MGo acknowledges support by MICCoM, as part of the Computational Materials Sciences Program funded by the U.S. Department of Energy, Office of Science, Basic Energy Sciences, Materials Sciences and Engineering Division.
SG, AMe, AV were supported by funding from the European Union’s Horizon 2020 research and innovation program under grant agreement No. 689868 (SOLSA).
AJM acknowledges networking support via the EPSRC Collaborative Computational Projects, CCP9 (EP/M022595/1) and CCP-NC (EP/T026642/1) and the EPSRC High-End Computing Consorium, UKCP (EP/P022561/1).
GMR acknowledges support from the F.R.S.-FNRS.
MSchei acknowledges support from the European Union's Horizon 2020 research and innovation program under the European Union's Grant agreement No. 676580 (NoMaD) and financial support from the Max Planck research network on big-data-driven materials science (BiGmax).
CO, CT, PC, MEs, FRo, SC acknowledge support by DOD-ONR (N00014-17-1-2090, N00014-17-1-2876) and  by the National Science Foundation under DMREF Grant No. DMR-1921909.
XY acknowledges support from the National Key Research and Development Program of China (Grants No. 2017YFB0701702 and 2016YFB0700501)

\section*{Author contributions statement}
All authors participated in discussions that lead to the OPTIMADE specification version 1.0, in particular, by taking part to the OPTIMADE workshops that took place at the Lorentz Center in Leiden (Netherlands) in 2016, and at the CECAM in Lausanne (Switzerland) in 2018, 2019, and 2020.
They also revised and approved the final version of the paper.\\
CWA, RAr, EB, GJC, SD, MLE, AF, AG, SG, AMe, FM, CO, GPi, GMR, MSchei, LT, CT, and DWi provided the key ideas for the design of the API, took part to the discussions about the issues opened on the Materials Consortia GitHub repositories, writing its specifications, and prepared the paper.\\
\noindent
CWA and MLE, with help from AF, SD and MSchei, write and maintain the optimade-python-tools library, which is based on initial work by DWi and MWu, with contributions from RAr, KC, AG, AMe, FM, LT and TP. This library forms the basis of the Materials Cloud, Materials Project, NOMAD and odbx providers as well as the client hosted on Materials Cloud.\\
\noindent
CD, GH, NM, GMR, and MSchef started the OPTIMADE workshop series, 
NM suggesting to dedicate these to the creation of a common API for databases of materials properties.
RAr, GJC, CD, SG, NM, GH, GPi, GMR, MSchef, MSchei, and CT were on the organizing committee. 
\noindent
CWA implemented the OPTIMADE server for the AiiDA framework and the Materials Cloud databases with contributions from GPi, LT, SK.\\
\noindent
RAr has contributed very actively to the OPTIMADE specification GitHub repository. He also implemented the OPTIMADE server in the high-throughput toolkit (\textit{httk}) open source framework used for the \textit{omdb} database.\\
\noindent
SD implemented the OPTIMADE server for the Materials Project provider, with assistance from MKH and PH. MKH wrote the OPTIMADE client in \textit{pymatgen}.\\
\noindent
MLE implemented the OPTIMADE server for the odbx provider, with assistance from AJM.\\
\noindent
AG and ML implemented OPTIMADE server for OQMD provider with assistance from VH
\noindent
SG, AMe, and LT contributed to the implementation of the OPTIMADE::Filter Perl library for the syntactical analysis of the OPTIMADE filter language.\\
\noindent
MSchei implemented the OPTIMADE server for the NOMAD provider with contributions from FM.\\
\noindent
CT, CO, PC, ME, FRo, and SC implemented the OPTIMADE server for the AFLOW provider.\\ 
\noindent
AV contributed to the long-term maintenance of the COD, to make it suitable for presentation via the OPTIMADE API.\\

\section*{Competing interests}

GJC is a shareholder and Director of Intellegens Ltd.
GH, GPe, GMR and DW are shareholders and Directors of Matgenix SRL.
The other authors declare no competing commercial interests.

\end{document}